\documentclass[paper,notoc]{JHEP3}
\usepackage{epsfig,cite}
\usepackage{amsbsy} 
\usepackage{epsfig}
\usepackage{graphicx}
\usepackage{amssymb,amsfonts,amsmath,bbold}

\title{Multi-gluon one-loop amplitudes numerically.}

\author{Achilleas Lazopoulos\\
  Institute for Theoretical Physics, ETH Zurich,\\
  8093 Zurich, Switzerland\\
  E-mail: \email{lazopoulos@itp.phys.ethz.ch}}

\abstract{ 
A {\tt c++} implementation of the $D_s$-dimensional unitarity cut  algorithm for the numerical evaluation of the virtual contribution to NLO QCD amplitudes is presented. The current version includes an arbitrary number of external gluons with gluonic propagators in the loop.  The building blocks are tree level color-ordered amplitudes with gluons and with gluons and two scalars in five dimensions. Numerical stability issues are addressed and agreement has been reached with the results published in the literature. } 

\newcommand{\be}{\begin{equation}}
\newcommand{\ee}{\end{equation}}

\newcommand{\bea}{\begin{eqnarray}}
\newcommand{\eea}{\end{eqnarray}}

\begin{document}

\section{Introduction}

In anticipation of the LHC data  the need for accurate calculations of Standard Model processes beyond the leading order is unequivocal~\cite{Bern:2008ef}. An algorithm that is able to perform such next to leading order calculations in a relatively generic and thus automatable way has been since long a withstanding goal of the community. Such a possibility seems within reach now, after the significant progress of the last two years, triggered by the OPP reduction method at the integrand level~\cite{Ossola:2006us},~\cite{Ossola:2007ax},~\cite{Ossola:2008xq}. This approach has been implemented in a publically released code, CutTools~\cite{Ossola:2007ax} and has been employed to calculate the six-photon amplitude~\cite{Ossola:2007bb} and the NLO QCD corrections to trivector boson production~\cite{Binoth:2008kt}. 

Based on the idea of numerical reduction at the integrand level, the authors of~\cite{Ellis:2008kd} employed unitarity-cut methods to recover the cut constructible part of the virtual NLO amplitude using tree level amplitudes as building blocks.  It was later shown~\cite{Giele:2008ve} that the rational part can also be evaluated numerically by applying the same algorithm in higher but integer number of dimensions, $D_s$. The generalization to fermions has been demonstrated in~\cite{Ellis:2008ir} and the first phenomenological applications using unitarity cuts have been achieved in~\cite{Ellis:2008qc}. 

In parallel, a semi-anaytic approach is engineered by the Black Hat collaboration~\cite{Berger:2008sj},\cite{Berger:2008sz},\cite{Berger:2008ag},\cite{Berger:2006cz},\cite{Berger:2006ci} 
in which the cut-constructible part of the amplitude is evaluated through a particular  complex-valued parametrization of the loop 
momentum. Thanks to the functional dependence on the complex parameter involved, one is able to separate integral contributions to a given cut in combination with OPP subtraction. The rational term is evaluated via a loop-level on-shell recursion relation.

An implementation of the $D_s$-dimensional unitarity cut algorithm~\cite{Giele:2008ve} (which we will call the EGKM algorithm in what follows) in {\tt C++} is presented in this paper, very similar to that of~\cite{Giele:2008bc} (written in {\tt FORTRAN 95}). The goal here is to build up a tool as generic as possible that can deal with processes of many external particles in an arbitrary theory with fermions, gauge bosons and scalars. When combined with an automated treatment of the real radiation (see~\cite{Frederix:2008hu},~\cite{Seymour:2008mu}) this completes the task of evaluating NLO corrections to tree-level matrix elements and thus can be used to upgrade to NLO existing matrix element generators.  The present paper announces the first step in this process, a program that calculates numerically loop amplitudes with gluons.

\section{The EGKM approach to loop amplitudes}
\label{the algorithm}
The implementation follows the algorithm of~\cite{Giele:2008ve} which is presented below in short. The virtual contribution to  QCD amplitudes involving one loop diagrams can be decomposed into a sum of terms in each of which color information appears as a factor multiplying a gauge invariant quantity called color-ordered amplitude. In the case of N-gluon amplitudes the color-ordered amplitudes are amplitudes with a fixed ordering of external legs including either a gluon or a fermion loop. In this paper only color-ordered amplitudes with gluon loops are considered. We then have
\be
A_{N,full}=\sum_{c=1}^{[N/2]+1}\sum_{\sigma \in S_N/S_{N;c}}Gr_{N;c}(\sigma)A_{N;c}
\ee
where
\be
Gr_{N;c}(\sigma)=Tr(t^{a_{\sigma(1)} } \ldots t^{a_{\sigma(c-1)}   }) Tr(t^{a_{\sigma(c)} } \ldots t^{a_{\sigma(N)}   })
\ee
\be
Gr_{N;1}=N_cTr(t^{a_{\sigma(1)} } \ldots t^{a_{\sigma(N)}   })
\ee
and $\sigma(i)$ is some permutation of $\{1\ldots N\}$, $S_{N;c}$ is the subset of $S_N$ that leaves the ordering indicated by the double trace invariant. Given that all $A_{N;c}$ for $c>1$ can be written as linear combinations of $A_{N,1}$, we will focus on the calculation of these in what follows.

A generic color-ordered one loop amplitude with N gluons is just a sum of the relevant feynman diagrams. Writing all Feynman diagram contributions under one denominator we have 
\be
A^{D_s}(p_1,p_2,\ldots,p_N)=\int [dl] \frac{N^{D_s}(p_1,p_2,\ldots,p_N;l)}{d_1d_2\ldots d_N}
\ee 
where
\be
d_1=l^2\;\;\;d_{i>1}=(l+p_1+p_2+\ldots+p_{i-1})^2
\ee
In the above expression $D_s$ denotes the dimensionality of the internal, unobservable, particles (gluons in our case). It depends on the regularization scheme and is set to $4$ in the Four Dimensional Helicity scheme (FDH)~\cite{Bern:2002zk} or to $4-2\epsilon$ in the 't Hooft - Veltman scheme (HV)~\cite{'tHooft:1972fi}. It is in general different than the dimensionality of the loop momentum (and that of the integral), which we denote by $D$. 

Since the external particles are kept strictly four dimensional, the dependence of the numerator on $D_s$ is linear for one loop, color-ordered, pure gluonic amplitudes, as can be seen by simple numerator algebra: $D_s$ enters only  as $g_{\mu\nu}g^{\mu\nu}=D_s$. 

Seen as a function of $D_s$, 

\be
A^{D_s}=A^0+A^1\cdot D_s
\ee

One can, therefore, for a given phase space point and for given external helicities, numerically evaluate $A^{D_s}$ for two values of $D_s$, determine the $D_s$ independent $A^0,A^1$ and thus get the full $A^{D_s}$ for arbitrary $D_s$. Thereafter, setting $D_s$ to $4$ recovers the FDH scheme  and to $4-2\epsilon$ the 't Hooft scheme. During all this the dimensionality of the loop momentum is kept to arbitrary $D$ with the constraint $D<D_s$. Only after the reduction is performed one can set $D\rightarrow 4-2\epsilon$ and evaluate the master integrals as a series in $\epsilon$. 

The problem of numerically evaluating full one loop amplitudes (including the rational part) reduces then to the problem of evaluating $A^{D_s}$ for two values of $D_s>D$. In the case only gluons are present, one can set $D_s=5,6$. In the following the calculation of $A^{D_s}$ for fixed $D_s$ is described.

It is well known that purely four-dimensional amplitudes can be reconstructed as linear combinations of scalar (i.e. with unit numerator) boxes, triangles, bubbles and tadpoles (higher scalar integrals are always reducible to the former \cite{vanNeerven:1983vr}):
\bea
A^{D_s=4}(p_1,p_2,\ldots,p_N)
&=&\sum_{Q=\{i_1,i_2,i_3,i_4\}} {d_{Q}\int[dl]\frac{1}{d_{i_1}d_{i_2}d_{i_3}d_{i_4}}}
      +\sum_{T=\{i_1,i_2,i_3\}}{c_{T}\int[dl]\frac{1}{d_{i_1}d_{i_2}d_{i_3}}}\nonumber\\
& &+\sum_{B=\{i_1,i_2\}}{b_{B}\int[dl]\frac{1}{d_{i_1}d_{i_2}}}
      +\sum_{S=\{i_1\}}{a_{S}\int[dl]\frac{1}{d_{i_1}}}
\label{reduction4}
\eea
with the summation extending to all possible combinations of propagators (with $i_j$ ordered thanks to the ordering of external legs). Here we use $Q,T,B,S$ as  collective indices representing ${i_1,i_2,\ldots}$ to keep the formulas readable.
 
Equation \ref{reduction4} is the result of a reduction process which can be seen as bringing the amplitude in the form
\bea
A^{D_s=4}(p_1,p_2,\ldots,p_N)&=&
\sum_{Q={i_1,i_2,i_3,i_4}}{  \int[dl]\frac{\bar{d}_{Q}(l)}{d_{i_1}d_{i_2}d_{i_3}d_{i_4}}}
+\sum_{T={i_1,i_2,i_3}}     { \int[dl]\frac{\bar{c}_{T}(l)}{d_{i_1}d_{i_2}d_{i_3}}} \nonumber \\
& &+\sum_{B={i_1,i_2}}     { \int[dl]\frac{\bar{b}_{B}(l)}{d_{i_1}d_{i_2}}}
+\sum_{S={i_1}}                  { \int[dl]\frac{\bar{a}_{S}(l)}{d_{i_1}}}
\eea
and then performing the loop integral.

Note here that the functions $\bar{d}_{Q}(l)$,$\bar{c}_{T}(l)$,$\bar{b}_{B}(l)$,$\bar{a}_S(l)$ depend on $l$ in a particular way. For example $\bar{d}_{Q}(l)$ consists of a piece independent of $l$ and a piece that is proportional to the four-dimensional subspace orthogonal to the one spanned by the three momenta that enter the propagators. The former is equal to $d_Q$ (the coefficient of the scalar integral in eq.~\ref{reduction4}) and the latter vanishes upon integration. Terms involving the components of $l$ in the subspace spanned by the momenta entering the propagators can be combined in propagators and hence appear as constant terms in $\bar{d}_{Q}(l)$ or  in one of  $\bar{c}_{T}(l)$'s. 
 
When one extends to arbitrary $D_s>4$ one can also have pentagons

\bea
A^{D_s}(p_1,p_2,\ldots,p_N)&=&\sum_i{ \int[dl] \frac{\bar{e}_E(l)}{d_{i_1}d_{i_2}d_{i_3}d_{i_4}d_{i_5}}}
+\sum_i{  \int[dl]\frac{\bar{d}_Q(l)}{d_{i_1}d_{i_2}d_{i_3}d_{i_4}}} \nonumber \\
& &+\sum_i{ \int[dl]\frac{\bar{c}_T(l)}{d_{i_1}d_{i_2}d_{i_3}}}
+\sum_i{ \int[dl]\frac{\bar{b}_B(l)}{d_{i_1}d_{i_2}}}
+\sum_i{ \int[dl]\frac{\bar{a}_S(l)}{d_{i_1}}}
\eea

The equivalent equation for the \textbf{integrands} of the two sides would be

\bea
A^{D_s}(p_1,p_2,\ldots,p_N;l)&=&\sum_i{  \frac{\bar{e}_E(l)}{d_{i_1}d_{i_2}d_{i_3}d_{i_4}d_{i_5}}}
+\sum_i{  \frac{\bar{d}_Q(l)}{d_{i_1}d_{i_2}d_{i_3}d_{i_4}}} \nonumber \\
& &+\sum_i{ \frac{\bar{c}_T(l)}{d_{i_1}d_{i_2}d_{i_3}}}
+\sum_i{ \frac{\bar{b}_B(l)}{d_{i_1}d_{i_2}}}
+\sum_i{ \frac{\bar{a}_S(l)}{d_{i_1}}}
\label{central}
\eea

\subsection*{pen-tuple cuts}

At this point one can multiply both sides of the above equation with five $d_i$'s. Then one sets the loop momentum such that these five propagators vanish (i.e. multiplies by a suitable $\delta$-function) and integrates over the loop momentum. 

The left hand side becomes then a product of tree-level amplitudes since the $\delta$-function sets the corresponding internal legs on-shell.
\bea
{\cal X}^{D_s}_{\alpha\beta\gamma\delta\epsilon}(\hat{l})&\equiv& \int [dl] A^{D_s}(p_1,p_2,\ldots,p_N;l)\; d_{\alpha} d_{\beta} d_{\gamma} d_{\delta} d_{\epsilon} \;\delta^D(l-\hat{l})\nonumber\\
&=&\sum_{\lambda_1,\ldots,\lambda_5} M^{D_s,\lambda_1\lambda_2}_{\alpha\beta}
M^{D_s,\lambda_2\lambda_3}_{\beta\gamma}
M^{D_s,\lambda_3\lambda_4}_{\gamma\delta}
M^{D_s,\lambda_4\lambda_5}_{\delta\epsilon}
M^{D_s,\lambda_5\lambda_1}_{\epsilon\alpha}
\label{lhs}
\eea
with $\hat{l}$ being the particular D-vector that makes the propagators vanish, and
\be
M^{D_s,\lambda_2\lambda_3}_{km}=A^{D_s}_{tree}(\hat{l}^{\lambda_1}+P_{j,k-1},p_k^{h_k},p_{k+1}^{h_{k+1}},\ldots,p_{m-1}^{h_{m-1}},-(\hat{l}-P_{k,m-1})^{\lambda_2}    ))
\ee
where we have explicitly denoted the helicities of external particles, and
\be
P_{k,m}\equiv p_k+p_{k+1}+\ldots+p_m
\ee

The sums in eq.\ref{lhs} extend over all polarization states of the internal, loop propagators that have now been put on-shell and have become external legs. Note that the momenta $\hat{l}$ are complex (which is why the three-particle tree-level amplitudes do not vanish). It is evident that this step can only be performed if the internal legs have integer dimensionality, so $D_s$ should be integer and larger than $4$. In the case of fermions running in the loop one needs also $D_s$ to be even.

Each of the contributing tree-level amplitudes can be evaluated numerically, thus yielding a value for the left hand side (eq.\ref{lhs}).

All terms in the right hand side vanish except the one of the relevant pentagon:
\be
\int [dl] \sum_E{  \frac{\bar{e}_E(l)}{d_{i_1}d_{i_2}d_{i_3}d_{i_4}d_{i_5}}}\; d_{\alpha} d_{\beta} d_{\gamma} d_{\delta} d_{\epsilon} \;\delta^D(l-\hat{l}) = \bar{e}_{\alpha\beta\gamma\delta\epsilon}(\hat{l})
\ee
so
\be
{\cal X}^{D_s}_{\alpha\beta\gamma\delta\epsilon}(\hat{l})=\bar{e}_{\alpha\beta\gamma\delta\epsilon}(\hat{l})
\label{pentagon_cut}
\ee

If one knows the functional form of $\bar{e}_{\alpha\beta\gamma\delta\epsilon}(l)$ and one has the freedom of finding many different $\hat{l}$'s that satisfy the unitarity constraints, one can solve for the coefficients of $\bar{e}_{\alpha\beta\gamma\delta\epsilon}$  and hence recover it fully. 

Let's call $Q_i$,$i=1\ldots4$ the sum of the external momenta between the $i$'th and the $i+1$'th cut. The five cut propagators can be written as 
\be
d_{\alpha}=l^2\;\;d_{\beta}=(l+Q_1)^2\;\;d_{\gamma}=(l+Q_1+Q_2)^2\;\;d_{\delta}=(l+Q_1+Q_2+Q_3)^2\;\;d_{\epsilon}=(l+Q_1+Q_2+Q_3+Q_4)^2
\ee
Then the solution of the unitarity constraints $d_i=0$ can be found as follows. Find four  $v_i,i=1..4$ such that $Q_i\cdot v_j=d_{ij}$. This can be done in a straightforward way with the Van Neerven-Vermaseren algorithm (see~\cite{vanNeerven:1983vr},~\cite{Ellis:2008kd}). Then 
\be
\hat{l}^{\mu}=V^{\mu}+\sqrt{-V^2}n_5^{\mu}
\label{pentagon_loop}
\ee
with
\be
V^{\mu}=x_iv_i^{\mu}
\ee
and $n_5^{\mu}$ the unit vector in the fifth dimension (i.e. $n_5=[0,0,0,0,i]$).
This automatically satisfies $d_{\alpha}=0$ as long as $n_5^2=1$.
The $x_i$'s are determined from the other unitarity constaints: $d_{\beta,\gamma,\delta,\epsilon}=0$.

In the pentagon case $\bar{e}(l)=e_0$ is independent of $l$ (see the discussion at\cite{Giele:2008ve}), so one $\hat{l}$ value fully determines $\bar{e}_{\alpha\beta\gamma\delta\epsilon}$. 

One should then repeat the process for all pentagon cuts, recovering all $e_0$ coefficients. At that point, the pentagon contribution to $A^{D_s}$ can be evaluated by multiplying the pentagon coefficient with the corresponding scalar pentagon integral. One can do better than that, though, since the pentagon is easily reduced in boxes as explained in~\cite{Giele:2008ve}. Hence the pentagon coefficients will be added to the coefficients of the five corresponding boxes.

\subsection*{Quadruple cuts}
Next, one proceeds to the four-cuts, where a further complication is met: upon taking the residue in both sides of eq.\ref{central} one has, on the right hand side, contributions from pentagons that share the same four propagators with the current cut. These  contributions  have to be subtracted from the left hand side to give the particular $\bar{d}_{i_j}(l)$. 

\be
\bar{d}_{\alpha\beta\gamma\delta}(\hat{l})={\cal X}^{D_s}_{\alpha\beta\gamma\delta}(\hat{l}) - \sum_{\epsilon \neq \alpha\beta\gamma\delta } \frac{e_{\alpha\beta\gamma\delta\epsilon}(\hat{l})}{d_\epsilon(\hat{l})}
\label{box_cut}
\ee

Let's call $Q_i$,$i=1\ldots3$ the sum of the external momenta between the $i$'th and the $i+1$'th cut. The four cut propagators can be written as 
\be
d_{\alpha}=l^2\;\;d_{\beta}=(l+Q_1)^2\;\;d_{\gamma}=(l+Q_1+Q_2)^2\;\;d_{\delta}=(l+Q_1+Q_2+Q_3)^2
\ee
Then the solution of the unitarity constraints $d_i=0$ can be found as follows. Find three  $v_i,i=1..3$ such that $Q_i\cdot v_j=d_{ij}$ and $n_1^{\mu}$ such that $n_1\cdot v_i=n_1\cdot Q_i=0$. This vector $n_1$ lives in the 4D subspace that is orthogonal to $Q_{1,2,3}$. This can be done in a straightforward way with the Van Neerven-Vermaseren algorithm (see~\cite{vanNeerven:1983vr},~\cite{Ellis:2008kd}). Then 
\be
\hat{l}^{\mu}=V^{\mu}+a_1 n_1^{\mu}+a_5 n_5^{\mu}
\label{box_momentum}
\ee
with
\be
V^{\mu}=x_iv_i^{\mu}
\ee
and $n_5^{\mu}$ the unit vector in the fifth dimension (i.e. $n_5=[0,0,0,0,i]$). The constraint $d_{\alpha}=0$ translates into \be
a_1^2+a_5^2+V^2=0
\label{quad_constraint}
\ee
The other three constraints are used to determine the numerical values of $x_i$.

The functional dependence of the box coefficient on $l$ is, 
\be
\bar{d}=d_0+d_1s_1+d_2s_e^2+d_3s_1s_e^2+d_4s_e^4
\ee   
with
\be
s_e^2=\sum_{i=4..D_s}(l\cdot n_i)^2\;\;\;\;s_1\equiv\cdot n_1
\ee
being the square of the projection of the loop momentum in the extra dimensions.

To determine the $d_i$'s we start with four dimensional loop momenta embedded in a $D_s$ dimensional space. Then 
to find $d_0$ and $d_1$ we need two (four-dimensional) values for $\hat{l}$.

Setting $a_5$ to zero we can get two different $\hat{l}$'s:
\be
\hat{l}^{\mu}_{1,2}=\hat{l}^{\mu}=V^{\mu}\pm \sqrt{-V^2} n_1^{\mu}
\label{box_loop}
\ee
from which $d_{0,1}$ are determined. 

Next, we set 
\bea
l^{\mu}_3: & a_1=\sqrt{-V^2}=a_5\\
l^{\mu}_4:&a_1=-\sqrt{-V^2}=-a_5\\
l^{\mu}_5:&a_1=\sqrt{-3V^2 /4}\;\;a_5=\sqrt{-V^2/4}
\eea 
for the three five-dimensional $\hat{l}_{3,4,5}$ that will determine $d_{2,3,4}$. Obviously any combination of $a_1,a_5$ with eq.\ref{quad_constraint} satisfied would be equally appropriate.

\subsection*{Triple cuts}
Similarly, for the triple cuts we have
\be
\bar{c}_{\alpha\beta\gamma}(\hat{l})={\cal X}^{D_s}_{\alpha\beta\gamma}(\hat{l}) - \sum_{\delta,\epsilon \neq \alpha\beta\gamma } \frac{\bar{e}_{\alpha\beta\gamma\delta\epsilon}(\hat{l})}{d_\epsilon(\hat{l})d_{\delta}(\hat{l})}
- \sum_{\delta \neq \alpha\beta\gamma} \frac{\bar{d}_{\alpha\beta\gamma\delta}(\hat{l})}{d_{\delta}(\hat{l})}
\label{triangle_cut}
\ee

Calling  $Q_i$,$i=1\ldots2$ the sum of the external momenta between the $i$'th and the $i+1$'th cut. The three cut propagators can be written as 
\be
d_{\alpha}=l^2\;\;d_{\beta}=(l+Q_1)^2\;\;d_{\gamma}=(l+Q_1+Q_2)^2
\ee
Then the solution of the unitarity constraints $d_i=0$ can be found as follows. Find two  $v_i,i=1..2$ such that $Q_i\cdot v_j=d_{ij}$ and $n_1^{\mu},n_2^{\mu}$ such that $n_i\cdot v_j=n_i\cdot Q_i=0$. Moreover we now demand $n_1\cdot n_2=0$. These vectors $n_1,n_2$ span the subspace in 4-d that is orthogonal to $Q_{1,2}$. Then 
\be
\hat{l}^{\mu}=V^{\mu}+a_1 n_1^{\mu}+a_2 n_2^{\mu}+a_5 n_5^{\mu}
\label{box_momentum}
\ee
with
\be
V^{\mu}=x_iv_i^{\mu}
\ee
and $n_5^{\mu}$ the unit vector in the fifth dimension (i.e. $n_5=[0,0,0,0,i]$). The constraint $d_{\alpha}=0$ translates into \be
a_1^2+a_2^2+a_5^2+V^2=0
\label{triple_constraint}
\ee

The functional form for $\bar{c}(l)$ can be chosen to be 
\be
\bar{c}(l)=c_0+c_1 s_1+c_2 s_2 + c_3(s_1^2-s_2^2)+s_1s_2(c_4+c_5s_1+c_6s_2) + c_7 s_1s_e^2+c_8s_2s_e^2+c_9s_e^4
\ee
where 
\be
s_i=l\cdot n_i
\ee
with $i=1,2$ and
$n_1$ being the unit vector in the tangent space determined by the momenta that enter the triangle vertices. Similarly
\be
s_e^2=\sum_{i=4..D_s}(l\cdot n_i)^2
\ee
is the square of the projection of the loop momentum in the extra dimensions.

To evaluate the first seven coefficients we restrict $\hat{l}$ to four dimensions. We need seven different $\hat{l}$'s conveniently chosen, and they are of the form\footnote{The following construction employs a discrete Fourier transform and was employed in\cite{Mastrolia:2008jb},\cite{Berger:2008sj} as an optimized way of constructing and solving the resulting system of equations.} 

\be
l_m^{\mu}=V^{\mu}+a_{1,m}n_1^{\mu}+a_{2,m}n_2^{\mu}
\label{triangle_loop}
\ee

with
\be
a_1=Rcos(\frac{2\pi m}{7})\;\;\;s_2=Rsin(\frac{2\pi m}{7})\;\;\;\; m=-3..3
\ee

\be
R\equiv\sqrt{ -V^2}
\ee

This setup leads to 
\be
\bar{c}(l_m)=\sum_{k=-3}^3 \lambda_k e^{i2\pi m/7}\;\;\;\;m=-3..3
\ee

where $\lambda_k$ are simple linear combinations of $c_i$. The system can be readily inverted to yield
\be
\lambda_k=\frac{1}{7}\sum_{-3}^3 e^{-i2\pi  §mk/7}\bar{c}(l_m)
\ee

from which $c_{0,\ldots,6}$ are easily obtained.

Next we set 
\bea
a1=0&a_2=\sqrt{-V^2/2}&a5=\sqrt{-V^2/2}\nonumber\\
a1=0&a_2=-\sqrt{-V^2/2}&a5=\sqrt{-V^2/2}\nonumber\\
a1=\sqrt{-V^2/2}&a_2=0&a5=\sqrt{-V^2/2}
\eea
Similarly to the quadruple cut case, any combination that satisfies eq.\ref{triple_constraint} is equally apropriate.

In case $|R|$ is approaching zero the above definitions become numerically problematic and one has to resort to some alternative setup.

\subsection*{Double cuts}
For double cuts we have 
\be
\bar{b}_{\alpha\beta}(\hat{l})={\cal X}^{D_s}_{\alpha\beta}(\hat{l}) - \sum_{\gamma\delta,\epsilon \neq \alpha\beta } \frac{\bar{e}_{\alpha\beta\gamma\delta\epsilon}(\hat{l})}{d_\epsilon(\hat{l})d_{\delta}(\hat{l})d_{\gamma}(\hat{l})}
- \sum_{\gamma\delta \neq \alpha\beta} \frac{\bar{d}_{\alpha\beta\gamma\delta}(\hat{l})}{d_{\delta}(\hat{l})d_{\gamma}(\hat{l})}
- \sum_{\gamma \neq \alpha\beta} \frac{\bar{c}_{\alpha\beta\gamma}(\hat{l})}{d_{\gamma}(\hat{l})}
\label{bubble_cut}
\ee

We now have one external momentum $Q_1$ and therefore one $v_1^{\mu}$ and three vectors in the orthogonal space, $n_{1}^{\mu}$,$n_2^{\mu}$,$n_3^{\mu}$. Hence,
\be
l^{\mu}=V^{\mu}+a_{1}n_1^{\mu}+a_{2}n_2^{\mu}+a_{3}n_3^{\mu}+a_{5}n_5^{\mu}
\ee
with
\be
a_1^2+a_2^2+a_3^2+a_5^2=0
\label{double_constraint}
\ee

The functional form of $\bar{b}(l)$ is (the numerator can contain up to two powers of $l^{\mu}$)
\be
\bar{b}(l)=b_0+b_1s_1+b_2s_2+b_3s_3+b_4(s_1^2-s_3^2)+b_5(s_2^2-s_3^2)+b_6s_1s_2+b_7s_1s_3+b_8s_2s_3+b_9s_e^2
\ee
with
\be
s_i\equiv l\cdot n_i
\ee
Restricting ourselves to four-dimensional loop momenta we need nine independent $\hat{l}_m$'s. It is, however, convenient to enlarge the number of equations to ten, so that we can use the inverse Fourier transform trick.
\be
l_m^{\mu}=V^{\mu}+a_{1,m}n_1^{\mu}+a_{2,m}n_2^{\mu}+a_{3,m}n_3^{\mu}
\label{bubble_loop}
\ee

with
\be
a_1=0\;\;\;a_2=Rcos(\frac{2\pi m}{5})\;\;\;a_3=Rsin(\frac{2\pi m}{5})\;\;\;\; m=-2..2
\ee
and
\be
a_1=Rcos(\frac{2\pi m}{5})\;\;\;a_2=0\;\;\;a_3=Rsin(\frac{2\pi m}{5})\;\;\;\; m=-2..2
\ee
with
\be
R^2\equiv -V^2
\ee

The last coefficient is evaluated by defining a five dimensional loop momentum as:
\be
a_1=\sqrt{-V^2/2}\;\;\;a_2=0\;\;\;a_3=0\;\;\;a_5=\sqrt{-V^2/2}
\ee
Similarly to the quadruple and triple cut cases, any combination that satisfies eq.\ref{double_constraint} is equally apropriate.

Note that double cuts that isolate a single, external, on-shell line would provide the coefficients of massless bubbles with massless, on-shell incoming momentum that vanish in dimensional regularization. Hence such cuts need not be considered.

\subsection*{Single cuts}
Single cuts are not needed in the pure gluonic case (they would multiply gluon tadpole scalar integrals which are zero in dimensional regularization).

\subsection{Master integrals}
Having evaluated all the coefficients of eq.~\ref{central} for the given phase-space point, we can now proceed to perform the loop integration. The constant parts of all coefficients $e_{E,0}$,$d_{Q,0}$,$c_{T,0}$,$b_{B,0}$ will multiply the corresponding master integrals. The latter are evaluated numerically. The sum of all these terms is equal to the cut-constructible part of the loop amplitude. 
\be
A_{cc}=\sum_{Q}\tilde{d}_{Q,0}I_Q+\sum_Tc_{T,0}I_T+\sum_B b_{B,0}I_B
\ee
where $\tilde{d}_{Q,0}$ includes $d_{Q,0}$ as well as contributions from reduced pentagons.
Any coefficient that depends on $n_i\cdot l$ will vanish upon integration over $[dl]$. There are, however, coefficients that depend only on $s_e^2$, namely  $d_2$,$d_4$,$c_9$ and $b_9$ of every cut. Those don't vanish upon integration. They contribute to the rational part of the loop amplitude.  These contributions can be reduced to (see section IV.D of~\cite{Ellis:2008kd}) to 
\be
A_R=-\sum_{Q}\frac{d_{Q,4}}{6}-\sum_{T}\frac{c_{T,9}}{2}-\sum_{B}\frac{b_{B,9}}{6}
\ee   
Note that $d_{Q,2}$ doesn't contribute to order $0$ in $\epsilon$.

The scalar integrals needed are all known analytically and have been implemented in libraries, first by G. J. van Oldenborgh  in FF written in {\tt FORTRAN 77} and then by Ellis and Zanderighi in QCDLoop written in {FORTRAN 90} \cite{Ellis:2007qk}. For massless theories, there is also the implementation\cite{vanHameren:2005ed}. We currently use the QCDLoop library with a wrap code for {\tt c++} but we plan to interface with the latter as well in the near future.

\section{Summary of the algorithm}
In this section we summarize the reduction algorithm for calculating numerically the virtual amplitude for a particular color-ordered set of gluons with fixed helicities in an integer $D_s$ number of dimensions. One should perform the reduction twice, for two values of $D_s$. In the case of gluons $D_s=5,6$, but see section~\ref{tree-level} on a note for $D_s=6$. 
\begin{itemize}
\item Find all the possible pent-uple cuts for the given color-ordered amplitude.
\item for each pent-uple cut
\begin{itemize}
\item evaluate the loop momentum that solves the unitarity constraints (eq.\ref{pentagon_loop}).
\item evaluate the amplitude residue as a product of tree level on-shell amplitudes with complex momenta (eq.\ref{pentagon_cut} - involves assigning polarization vectors to the cut legs that carry complex momentum).
\item store the coefficient of the cut.
\item reduce the pentagon to boxes and assign the coefficients accordingly.
\end{itemize}
\item find all possible quadruple cuts.
\item for each quadruple cut
\begin{itemize}
\item find the set of loop momenta that solve the unitarity constraints and are convenient for solving the resulting system (eq.\ref{box_loop}).
\item evaluate the amplitude residue as a product of four tree level amplitudes. 
\item subtract the pentagon contributions evaluated at the particular $\hat{l}$ (eq.\ref{box_cut}).
\item solve the system of equations for $d_{0,\ldots,4}$.
\item store the value of the cut.
\end{itemize}
\item find all possible  triple cuts.
\item for each triple cut
\begin{itemize}
\item find the set of loop momenta that solve the unitarity constraints and are convenient for solving the resulting system (eq.\ref{triangle_loop}).
\item evaluate the amplitude residue as a product of three tree level amplitudes.
\item subtract the pentagon and box contributions evaluated at the particular $\hat{l}$ (eq.\ref{triangle_cut}).
\item solve the system of equations for $c_{0,\ldots,9}$.
\item store the value of the cut.
\end{itemize}

\item find all possible double cuts.
\item for each double cut
\begin{itemize}
\item find the set of loop momenta that solve the unitarity constraints and are convenient for solving the resulting system (eq.\ref{bubble_loop}).
\item evaluate the amplitude residue as a product of two tree level amplitudes.
\item subtract the pentagon, box and triangle contributions evaluated at the particular $\hat{l}$ (eq.\ref{bubble_cut}).
\item solve the system of equations for $b_{0,\ldots,9}$.
\item store the value of the cut.
\end{itemize}

\item multiply the relevant coefficients with the master integrals to get the final result
\end{itemize}

\subsection{Polarization vectors}
From the discussion in the previous section it is evident that one needs polarization vectors for $D_s=5,6$ four- and five-dimensional loop momenta. There are $D_s-2$ polarization vectors in each case.

When the loop momentum is itself four-dimensional one can use the traditional two polarization vectors perpendicular to it and $D_s-4$ polarization vectors that are unit vectors in the extra-dimensional subspace.  Care has to be taken, however, since the loop momentum is  also complex. The only requirement for choosing polarization vectors for the internal loop momentum is that they satisfy the proper polarization sums. We employ polarization vectors defined as spinor products which have polarization sums corresponding to the axial gauge, in that case. 

When the loop momentum is five-dimensional there are three polarization states. The corresponding vectors can be defined in a way analogous to the polarization vectors of a massive gauge boson in four dimensions, since then
\be
l^{\mu}=l_{4D}^{\mu}+\tilde{l}^{\mu}\rightarrow l^2=l_{4D}^2-\tilde{l}^2=0
\ee 
The polarization vectors corresponding to $l^{\mu}_{4D}$ with mass $\tilde{l}^2$ are the same as those corresponding to $l^{\mu}$ in 5 dimensions.

\subsection{Tree-level amplitudes}
\label{tree-level}
Since one can avoid entirely a fully six-dimensional loop momentum (see the discussion on the necessary loop momenta of section~\ref{the algorithm}), it is easy to see that the six-dimensional tree-level amplitudes needed for the construction of the cut residues always decouple into five-dimensional tree-level amplitudes (when the polarization of the loop legs are $\pm,5$) or amplitudes with scalars in the place of the loop legs (when the loop legs are both polarized in the 6-th dimension). The tree-level amplitude can always be written as
\be
A^6(p_1,\epsilon_1;\ldots;p_n,\epsilon_n)=\epsilon_1^{\mu}M^6_{\mu\nu}\epsilon_n^{\nu}
\ee
Since $M^6_{\mu\nu}$ is purely (at most) five-dimensional\footnote{It is five-dimensional when the loop momentum is five dimensional, and four-dimensional otherwise.}, if $\epsilon_1$ and $\epsilon_n$ are polarized along the 6th dimension (i.e. they are equal to [$0,0,0,0,0,1$]) the only non-zero term will occur for 
\be
M^6_{\mu\nu}\rightarrow A^6_{sc}g_{\mu\nu}
\ee
where $A^6_{sc}$ is the part of the subamplitude proportional to $g_{\mu\nu}$. Therefore one really needs tree-level amplitudes with gluons in $D_s=5$ and tree-level amplitudes with gluons and scalars in $D_s=5$.

Tree-level color-ordered amplitudes with gluons can be calculated recursively using the Berends-Giele recursion relation ~\cite{Berends:1987me}. One needs look-up tables to avoid calculating the same subamplitude more than once. Alternatively, one can use a bottom-up Berends-Giele approach, as employed in ~\cite{Kanaki:2000ey} in which it is by construction guaranteed than no subamplitude is calculated more than once. We have experimented with the bottom-up approach with explicit code for each number of external legs generated and compiled once. It is slightly faster for $N\leq 12$ but becomes slower for $N>12$ presumably due to code bloating. Here we use the top-down apporach. The gluon-scalar amplitudes are similarly generated. 

Since the five-dimensional gluon and scalar color-ordered amplitudes are the building blocks of this algorithm, efficient evaluation of them is of paramount importance. In table~\ref{treetable} we give  the CPU time needed to evaluate a given tree-level color-ordered amplitude in the gluon and scalar case for a number of external legs varying from four to twentytwo. Typical cpu times needed for the calculation are estimated by averaging over the time needed for the evaluation of $10^5$ phase space points generated by ${\tt RAMBO}$\cite{Kleiss:1985gy} with a given set of cuts ($|n_i|<3$,$p_{T,i}>0.01\sqrt{s}$,$R_{i,j}\equiv\sqrt{\phi_{i,j}^2+n_{i,j}^2}>0.4$) on an Intel Xeon X5450 {\tt@3.00GHz}.

\vspace*{12px}
\begin{table}
\begin{center}
\begin{tabular}{|c|c||||c|c||||c|c|}
\hline
\multicolumn{6}{|c|}{Time for a tree amplitude}\\
\hline
\hline
N&t($\mu$s)&N&$t (\mu$s)&N&t($\mu$s)\\
\hline
4 & 5 & 11 & 178&18&	1326\\
5 & 9 & 12 & 252&19&	1649\\
6 & 17 & 13 & 351&20&	2032\\
7 & 29 & 14 & 475&21&	2482\\
8 & 50 & 15 & 631&22&	3004\\
9 & 79 & 16 & 818 &&	\\ 
10 & 121 & 17 & 1048&&	\\ 
\hline
\end{tabular}

\caption{CPU time averages for tree-level amplitudes over $10^5$ phase space points generated by ${\tt RAMBO}$\cite{Kleiss:1985gy} with a given set of cuts ($|n_i|<3$,$p_{T,i}>0.01\sqrt{s}$,$R_{i,j}\equiv\sqrt{\phi_{i,j}^2+n_{i,j}^2}>0.4$) on an Intel Xeon X5450 {\tt@3.00GHz}.\label{treetable}}
\end{center}

\end{table}  

\section{Numerical issues}

Issues of numerical stability have to be addressed in any algorithm performing numerical integration of  one-loop amplitudes, due to the possible appearance of vanishing (or almost vanishing) Gram determinants in one or the other part of the calculation. In this particular algorithm the problem can appear in the evaluation of the reduction coefficients, and, in particular, when calculating the solution to the unitarity constraints for a given cut. Moreover, the subtraction of the contribution of higher cuts in quadruple or lower cuts is another potential source of loss of accuracy. 

The way to deal with this problem employed here is to check for any given point, the coefficient of the poles in the $\epsilon$-expansion, known a priori in  an analytic form thanks to~\cite{Catani:1998bh},\cite{Giele:1991vf},\cite{Kunszt:1994mc}. If the  (normalized) difference between the calculated pole coefficient and the analytically evaluated one is more than a predetermined value (say $10^{-4}$) the phase-space point is considered problematic. Thereafter the whole algorithm is performed again using quadruple precision complex numbers (implemented in the QD library~\cite{QDlib}). The quadruple version of the code is of course much slower than the native double precision  version, so one hopes that the fraction of times it needs to be employed is relatively small. In the case of six gluons approximately $5\%$ of the  points need quadruple precision if the switch accuracy is set to $10^{-4}$.

\begin{figure}
\begin{minipage}[b]{0.5\linewidth} 
\centering
\includegraphics[angle=270,width=6cm]{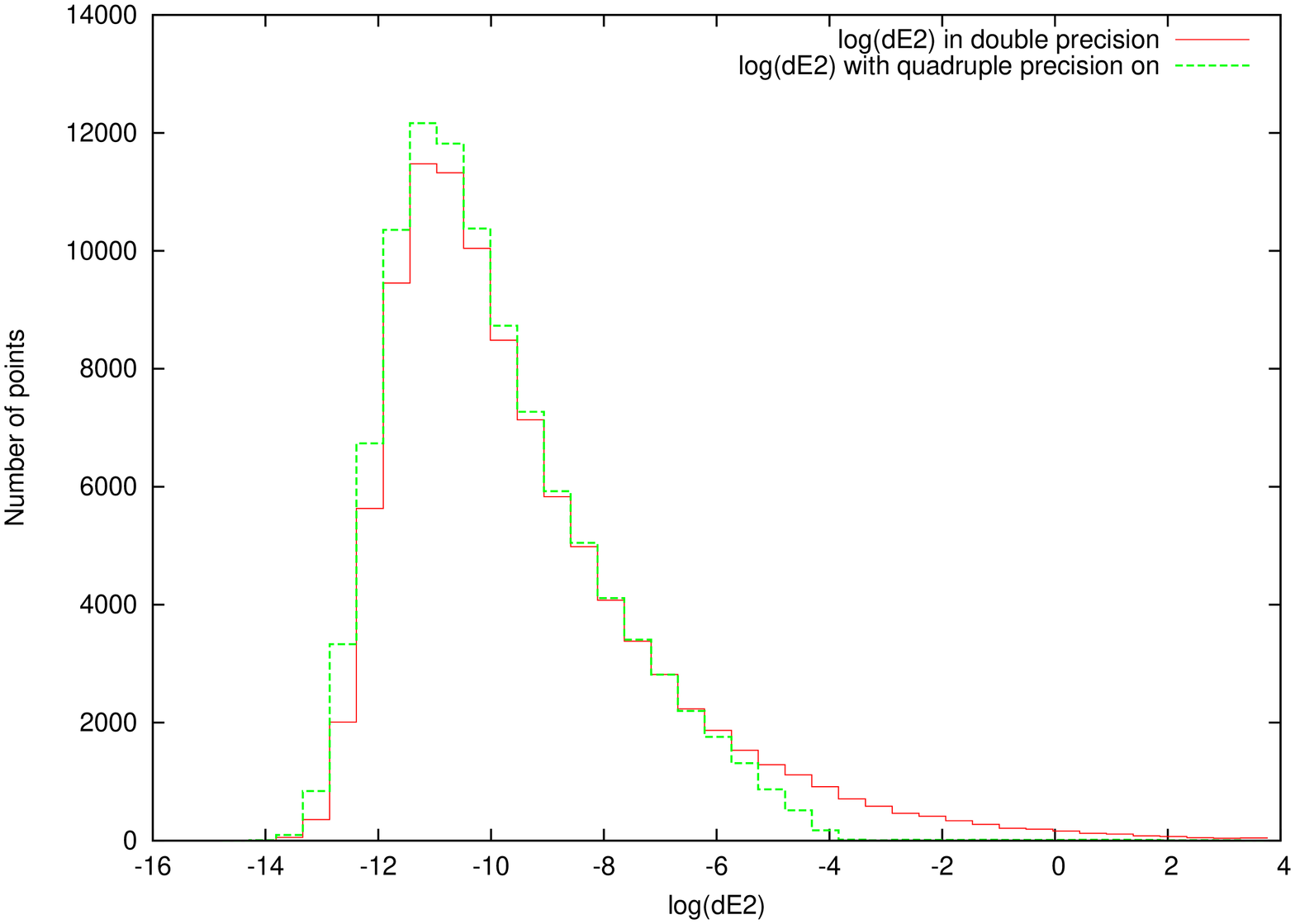}
\caption{The logarithm of the relative error for the double pole in $N=6$.}
\label{6paccuracy1}
\end{minipage}
\hspace{0.5cm} 
\begin{minipage}[b]{0.5\linewidth}
\centering
\includegraphics[angle=270,width=6cm]{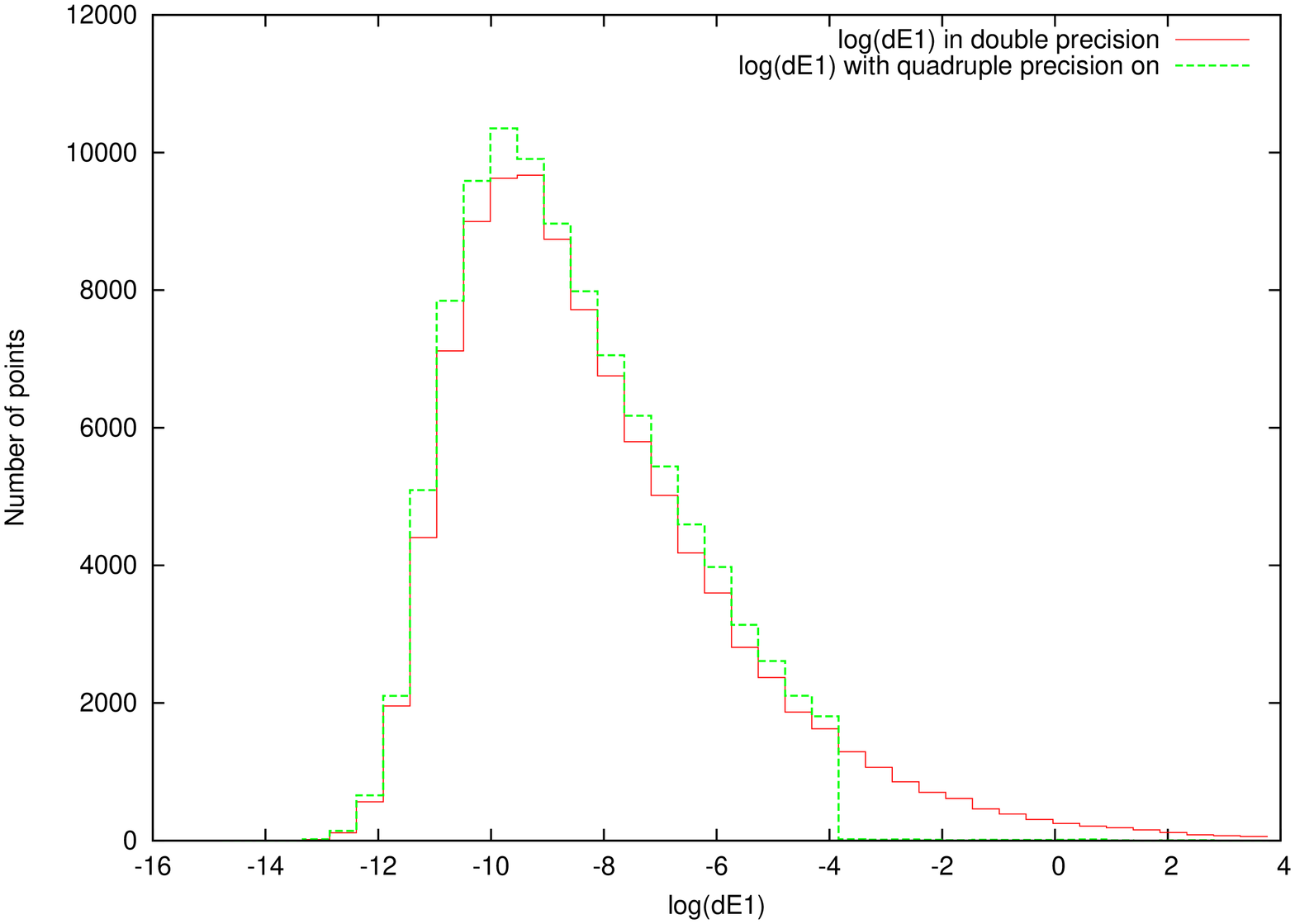}
\caption{The logarithm of the relative error for the single pole in $N=6$.}
\label{6paccuracy2}
\end{minipage}
\end{figure}

In figures~\ref{6paccuracy1},\ref{6paccuracy2} we show the log of the relative precision,
\be
dE_{1,2}=\frac{E_{1,2}-E_{analytic,1,2}}{E_{analytic,1,2}}  
\ee
for the double and single poles for the double precision version and for the version with the quadruple precision switched on. 
A similar distribution for the relative error is observed when the number of gluons increases, albeit the peak of the distribution shifts towards larger errors. This is expected to pose serious problems when $N>12$, but given the fact that one can even resort to arbitrary precision arithmetics (APREC~\cite{APREC}), cases of phenomenological importance should always be manageable with a relatively small penalty in time.

Note also that, despite the fact that relative errors for the poles are correlated with each other and with the relative error for the finite part, the latter is not guaranteed to behave well when the former do, due to potential instabilities in coefficients of finite scalar integrals.  There is a variety of checks one can perform to try to detect such cases so that quadruple precision arithmetics is switched on.

\section{Performance}

\begin{figure}[ht] 
   \centering
   \includegraphics[angle=270,width=130mm]{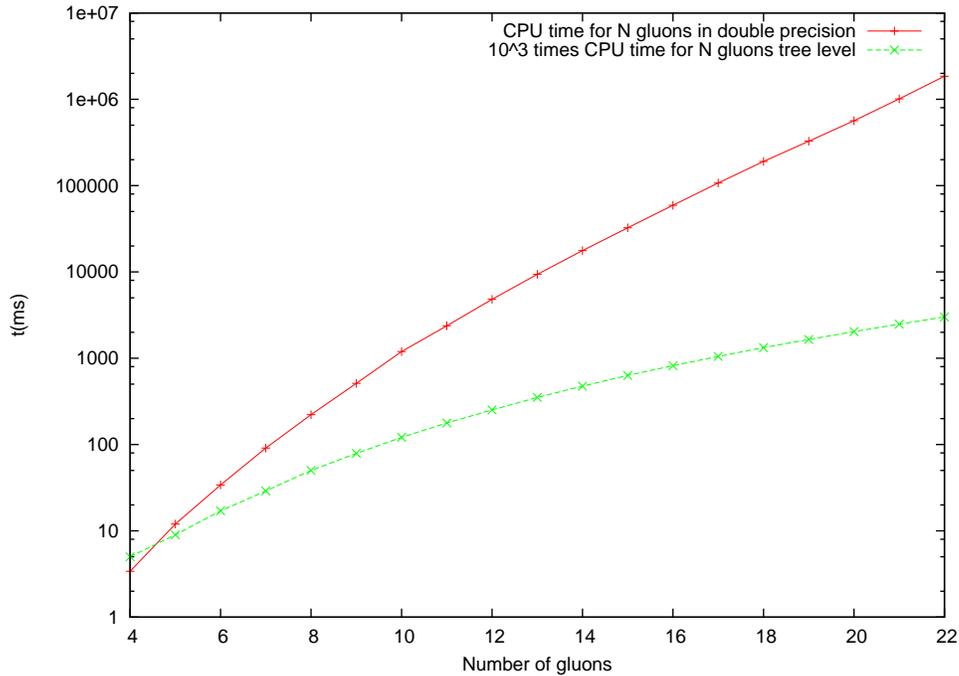} 
   \caption{CPU time for the color-ordered, helicity-fixed, N gluon amplitude in double precision on an Intel Xeon X5450 {\tt@3.00GHz} }
   \label{fig:speed}
\end{figure}

The code is presently able to calculate the full virtual part of color-ordered amplitudes with gluons, up to an arbitrary number of external legs and for arbitrary helicities, including the $1/e^2$ and $1/e$ coefficients, as well as the rational part of the finite term in the $\epsilon$-expansion. 

Checks with published results for specific phase-space points at~\cite{Giele:2008bc} and~\cite{Giele:2008ve} have been performed successfully.

Typical cpu times needed for the calculation are estimated by averaging over the time needed for the evaluation of $10^5$ phase space points generated by ${\tt RAMBO}$~\cite{Kleiss:1985gy} with a given set of cuts ($|n_i|<3$,$p_{T,i}>0.01\sqrt{s}$,$R_{i,j}\equiv\sqrt{\phi_{i,j}^2+n_{i,j}^2}>0.4$) on an Intel Xeon X5450 {\tt@3.00GHz}. Table~\ref{nlotable} shows those averages for $N=4..22$. In figure~\ref{fig:speed} these averages are plotted together with the equivalent tree-level ones multiplied by $10^3$.

\vspace*{12px}
\begin{table}
\center{
\begin{tabular}{|c|c|||c|c|||c|c|||c|c|}
\hline
N&t(ms)&N&t(ms)&N&t(ms)&N&t(ms)\\ \hline
4& 3.4&	9&512	&14&17675	&19	&327020\\ \hline
5&12&	10&1193	&15&32465	&20	&564330\\ \hline
6&34&	11&2376	&16&59145	&21	& 1011710\\ \hline
7&91&	12&4809	&17&107145	&22	& 1848280\\ \hline
8&222&	13&9380&18&190635	& &\\ \hline
\end{tabular}
\caption{Typical cpu times needed for the calculation are estimated by averaging over the time needed for the evaluation of $10^5$ phase space points generated by ${\tt RAMBO}$~\cite{Kleiss:1985gy} with a given set of cuts ($|n_i|<3$,$p_{T,i}>0.01\sqrt{s}$,$R_{i,j}\equiv\sqrt{\phi_{i,j}^2+n_{i,j}^2}>0.4$) on an Intel Xeon X5450 {\tt@3.00GHz}.\label{nlotable} }
}

\end{table}
\section{Outlook}
 
 We have presented here the first step in an effort to produce a generic implementation of the EGKM algorithm\cite{Giele:2008ve} for calculating numerically one-loop amplitudes. The current version involves gluons as external and as virtual particles. The necessary building blocks are color-ordered  multi-gluon tree-level amplitudes and tree-level amplitudes with gluons and scalars  in five dimensions. They are all evaluated via a Berends-Giele\cite{Berends:1987me} recursion relation. 
 
 The performance of the code is quite satisfactory and agreement has been reached with previously published results.  Numerical stability issues are all addressed by resorting to a quadruple precision version of the code. 
 
There are ample phenomenological applications for a generic code that includes fermions and electroweak gauge bosons, all within the immediate reach of the $D_s$-dimensional unitarity cut approach to OPP reduction, as demonstrated explicitly in~\cite{Ellis:2008qc}. Extending the implementation to (massless or massive) fermions and gauge bosons is relatively straightforward. To this generalization we turn in the future.   
  
\section*{Acknowledgments}
I would like to thank Kirill Melnikov for insightful discussions during the first half of this project. This work was supported by the US department of Energy under the contract DE-FG03-94ER-40833 and by the  Swiss National Science Foundation under the contract 200021-117873.

\bibliographystyle{is-unsrt}  
\bibliography{gluons}

\end{document}